\documentclass[pre,twocolumn,a4paper,floatfix]{revtex4}
\usepackage{amsmath}
\usepackage{graphicx}
\begin{document}

\title{Risk perception in epidemic modeling}
\author{Franco Bagnoli}
\affiliation{Department of Energy, University of Florence\\
Via S. Marta,3 I-50139 Firenze, Italy\\
also CSDC and INFN, sez.\ Firenze}
\email{franco.bagnoli@unifi.it}
\author{Pietro Li\`o}
\affiliation{Computer Laboratory, University of Cambridge\\
JJ Thompson Avenue, CB30FD Cambridge UK}
\author{Luca Sguanci}
\affiliation{Centro per lo Studio di Dinamiche Complesse (CSDC),
University of Florence, Via G. Sansone 1, Sesto Fiorentino (FI)}

\date{\today}

\begin{abstract}
  We investigate the effects of risk perception
  in a simple model of epidemic spreading.
  We assume that the perception of the risk of being infected 
  depends on the fraction of neighbors that are ill. 
  The effect of this factor is to decrease the
  infectivity, that therefore becomes a dynamical component of
  the model. We study the problem in the mean-field approximation and 
  by numerical simulations for regular, random and scale-free
  networks. 
  We show that for homogeneous and random networks, there
  is always a value of perception that stops the epidemics. In the
  ``worst-case'' scenario of a scale-free network with diverging input
  connectivity, a linear perception cannot stop the epidemics; however
  we show that a non-linear increase of the perception risk
  may lead to the extinction of the disease. This transition is 
  discontinuous, and is not predicted by the mean-field analysis. 
\end{abstract}


\maketitle

\section{Introduction}

In spring 2006, the potential threat of bird flu dominated headlines in
UK newspapers. On 26 March 2006 The Sun has called it ``the day we all
dreaded", while the Guardian says avian flu is ``almost certain to
spread to wild birds across the UK". The Daily Telegraph adds that the
most likely human victims will be poultry farmers, who will be
bankrupted. But the Mirror calls for calm, saying people have a better
chance of winning the lottery than catching the virus. Interestingly,
given a certain amount of clustering of wealth residents and
correlation between wealth and readers preference, this would translate
into a differently informed neighborhood. When the epidemic is over its
peak or other news has just peaked or media has ``cried wolf" too many
times over unfounded health scares, there is a quick drop in the
attention to that disease (something similar is reported nowadays for
HIV). In other parts of the world, for example Indonesia, a country
with 18000 islands, people reacted differently to the bird flu
epidemics. Despite awareness campaigns in the media and even
door-to-door visits in some of the islands, many Indonesians remained
oblivious to the dangers of being in contact with diseased birds, and
aware of the need to inform the authorities and implement a cull. Note
that awareness campaigns, such as during the SARS epidemics, are
expensive and may result in culling, reductions in commerce, travels
and tourism. The media hype over epidemics threat has a close
similarity in how worried or fatalist, resilient, skeptical or cheeky
may be friends and neighborhood. Therefore, the individual perception
of the risk of becoming infected is a key factor influencing the
spreading of an epidemics and, toward realistic inference,
epidemiological models should incorporate such
parameter~\cite{Vespignani1}.

In order to investigate the effect of risk perception in influencing
the spreading of a disease, let us start from simple, yet meaningful
models, such SIS or SIR ones.
These models are defined on a network where
individuals or groups of individuals corresponds to the
nodes and links represent social contacts and
relationships among them.  Most of classical studies used either
a regular lattice, or a random one.
Both of those choices are
characterized by a well defined value of the mean connectivity
$\langle k\rangle$, and small variance $\langle k^2\rangle-\langle
k\rangle^2$.
As shown by Watts and Strogatz~\cite{WS1998}, the simple rewiring of a
small fraction of links in an otherwise regular lattice results in a
sudden  lowering of the diameter of the
graph, without affecting the average connectivity or the degree of
clustering.  This \emph{small world} effect manifests itself
in a dramatic shortage of the distance between any two individuals,
almost without affecting the local perception of the network of
contacts.  The consequences for epidemics spreading are important: just
a few long-distance connections may promote the spreading of a disease
in rural areas, whereby an epidemic would otherwise diffuse very
slowly.

However, the investigations of social networks have shown that
they are quite different from regular and random
graphs~\cite{PV2001b,NewmanSIAM}.  The probability distribution of
contacts often
exhibits a power-law behavior ($P(k)\propto k^{-\gamma}$), with an
exponent $\gamma$ between two and
three~\cite{BarabasiAlbert,Dorogovtsev}. This distribution is
characterized by a relatively large number of highly connected hubs,
which are presumably responsible for epidemics spreading. Moreover,
such distributions have a diverging second moment $\langle k^2\rangle$
for $\gamma \le 3$ and a diverging average connectivity $\langle
k\rangle$ for $\gamma \le 2$.

The influence of the connectivity on the spreading dynamics is well
outlined by a simple mean-field analysis.
Let us consider for the moment a tree with fixed connectivity
$k$. In a SIS
model with immediate recovery dynamics, a single infected individual
may infect
up to $k$ neighbors~\cite{note}, each one with probability $\tau$.
The 
temporal behavior of the mean fraction $c$ of infected individuals is
given by
\begin{equation}\label{cm}
  c'= \sum_{s=1}^k \binom{k}{s} c^s(1-c)^{k-s}(1-(1-\tau)^s),
\end{equation}
where $c\equiv c(t)$,  $c'\equiv c(t+1)$ and the sum runs over the
number $s$ of infected individuals. The basic reproductive
ratio $R_0$~\cite{May} is simply given by $R_0 = k \tau$, so that the
epidemic
threshold $R_0=1$ corresponds to $\tau_c=1/k$. This means that for a
fixed connectivity,
only diseases with an infectivity less than $1/k$ do not spread. 

In heterogeneous networks (nodes with different connectivity) the
mean field analysis, reported in
Section~\ref{meanfield}, gives $\tau_c=\langle k^2\rangle/\langle
k\rangle$.  In the case $\langle
k^2\rangle\simeq \langle k\rangle^2$, $\tau_c$ is again equal to
$1/\langle k \rangle$. 
 
In summary, the result is that on very non homogeneous networks,
with diverging second moment 
$\langle k^2\rangle$ (and even worse on those with diverging average 
 connectivity $\langle k\rangle$), a
disease will always spread regardless of its intrinsic
morbidity~\cite{Boguna}. 

This result can be modified by the assortativity degree of the
network
and by the presence of loops, not considered in tyhe mean-field
analysis. In
networks with assortative connections (hubs are preferentially
connected to other hubs), it may happen that epidemics spread for
any finite infectivity even when the
second moment is not diverging~\cite{Vazquez,MorenoGomez}, while for
disassortative networks the reverse is true, epidemics may be stopped
by lowering the infectivity  with random vaccination campaigns, even
in the presence of a diverging second moment~\cite{Vazquez}. This is
particularly evident in networks lacking the small-world property
(consequence of high disassortativity)~\cite{VazquezBoguna,Eguiluz}.

In small-world networks with diverging second moment, it  is quite
difficult to stop an epidemics. The most common recipes
are vaccination campaigns (removal of nodes) or modification of the
social structure (removal of links), that mathematically corresponds
to site and bond percolation problems. To be efficient, a vaccination
camping must be targeted to hubs, either directly~\cite{PV2001b}
or implicitly, for instance by exploiting the fact
that hubs are the most probable neighbors of many nodes~\cite{Cohen}.

The modification of the social structure can be obtained by
cohercitive methods (quarantine, etc.) or by rising alerts so to
modify travelling and business patterns, but this option may be so
expensive
that the amount of money put into restoring the previous situation
may exceed that used to cure ill people~\cite{SARS}.

However, epidemics in modern world are relatively uncommon, and most
of
them are stopped quite easily in spite of the presence of high
network connectivity.
The existence of an epidemic threshold on
such networks has motivated the investigation of the effects of
connectivity-dependent infectivity~\cite{MN2002,Olinky,Callaway}. In
this
latter case, most of investigations have been performed using
mean-field techniques, thus disregarding the presence of loops.

Loops are irrelevant at and near the percolation
threshold~\cite{loops}, and therefore one can threat the network as a
tree in these conditions. However, for processes evolving on
percolating networks, this assumption may not hold.

At present, the basic models used
do not take into consideration the \emph{knowledge} that all
human beings
have nowadays about the mechanisms of diffusion of diseases. In
fact, even in the absence of vaccination campaigns, a disease that
manifests itself in a visible way induces modifications in the
social network: lower frequency of contacts (usage of mass
transportation systems), higher level of personal hygiene,
prevention measures (masks), etc. 
Indeed, recent works stress the importance of using a
time-dependent bare infectivity to
reproduce real patterns of
epidemics~\cite{RT2003,LS2003,Geisel,Kamo}. 

Viruses with high mutation
rates (like computer viruses) follow a dynamics which is more
similar to SIS than to SIR~\cite{PV2001}, even in the presence of
immunization. On
the other hand, the origin of vaccination come from
cross-immunization conferred by strains with lower pathogenicity.

We shall study here a SIS model in which the bare infectivity of
the spreading agent is modulated by a term that tries to model the
effects of the perception of the risk of being infected.

We assume that this perception is just  
an increasing function of the fraction of ill people in the
neighborhood, given that the illness presents visible symptoms.
This assumption is modeled after the
heuristic-systematic information-processing model~\cite{fear}, that
simply states that attitudes are formed and modified as people gain
information about a process. In the absence of explicit alarm or
communication, the only way of gaining this information is though
examination of people in the neighborhood. Individuals can process 
information in two ways: either heuristically, using simple an
semi-unconscious schemes, or carefully examining them in a rational
way. Investigations about the effects of advertisements, especially
those exploiting fear, show that the first way is rather common and
predictable, except in the limit of high level of fear, that may
induce repulsion towards the brand, or very low level, that may
trigger the reflexive mechanism and a more careful evaluation of the
message. 

In this work we simply assume that the local information (not
enhanced by alarms)
about the incidence of the illness translates into a lowering of the
infection probability, implementing only the ``linear part'' of
the information-processing model.
In principle, is possible to compare the
effective susceptibility to infection for diseases that manifest
themselves in a visible and in an invisible way and test
experimentally this hypothesis.
 
In our model, the infectivity is a dynamical quantity. Although
the idea of modulating the infectivity of the infection process is
not new, it is generally studied (mostly in the mean-field
approximation) as as a function  of
time~\cite{RT2003,LS2003,Geisel,Kamo}, of
connectivity~\cite{MN2002,Olinky} and/or depending on the total
infection level~\cite{Glendinning, Grenfell}. In this latter approach,
a nonlinear
growing dependence of the infection rate on the total number of
infected people may originate bifurcation and chaotic oscillations. 

As we shall show in the following, mean-field analysis may
not capture the essential phenomena in highly connected networks.
Moreover, we study the case of decreasing infection rate with
increasing local infection level, that might also induce chaotic
oscillations at the mean-field level (See Ref.~\cite{perpignan} and
Section~\ref{model}). However, one
should consider that chaotic oscillations on networks easily
desynchronize, and the resulting  ``microscopic chaos'' is quite
different from the synchronous oscillations predicted by mean-field
analysis~\cite{Boccara}, that may nevertheless be observed in
lattice models the presence of long-range coupling~\cite{longrange}.

We explicitly describe the model in Section~\ref{model}, analyze it
using mean-field techniques in Section~\ref{meanfield} and study
numerically its behavior on different types of networks in
Section~\ref{numerics}. Conclusions and perspectives are drawn in the
last section.

\section{The model}\label{model}
In this paper we study the dynamics of an infection spreading over
a network of $N$ individuals. We use different kinds of networks:
regular, with long-range rewiring~\cite{WS1998}, random and
scale-free~\cite{BarabasiAlbert}. The network structure is considered
not to depend on the infection level.

Let us
denote by $P(k)$ the probability distribution of connectivity $k$. We
shall denote by $z=\mu_1(P)$ the average connectivity (first moment of
the distribution), $z=\langle k \rangle = \sum_k k P(k)$, and by
$\mu_2\def \mu_2(P)=\langle k^2\rangle$ the second moment, $\mu_2 =
\sum_k k^2 P(k)$. 
In the case
of regular lattice with eventual rewiring, $P(k) = \delta_{kz}$ and
$\mu_2=z^2$. The rewiring of the network is performed by starting from
a regular lattice in one dimension, detaching a fraction $p$ of links
from one end and attaching them to randomly chosen nodes. The
regular case is studied numerically in one dimension. Simulations on
the rewired network are performed both in the quenched and in the
annealed cases.

For random
graphs, studied only at the mean-field level, the probability
distribution is assumed to be Poissonian, 
\[
P(k) = \frac{z^k e^{-z}}{k!},
\]
corresponding to drawing $Nz$ links at random among the $N$ nodes
($\mu_2=z$).

The scale-free network that we study numerically is asymmetric: each
node $i$  has a certain number $k_{\text{in}}(i)$ of
input contacts and $k_\text{out}(i)$ of output ones, and was grown
using the following rule.

We start with a ring of $K$ nodes, and we add the other $N-K$ nodes by
choosing, for each of them, $K$ connected nodes $j_n, n=1,\dots,K$,
with probability $k_\text{in}(j_n)/\sum_{l=1}^N k_\text{in}(l)$
(preferential attachment). The node being attached is added to the
\emph{inputs} of the chosen nodes. We also choose another node at
random and add it to the list of input nodes of the new node. This
process simulates the growing of a social network in which a new node
(a family or an individual) is born from another one (the ones that is
added as input of the newborn node) and joins the society according
with the popularity of nodes.

\begin{figure}[t]
\begin{center}
\includegraphics[width=\columnwidth]
{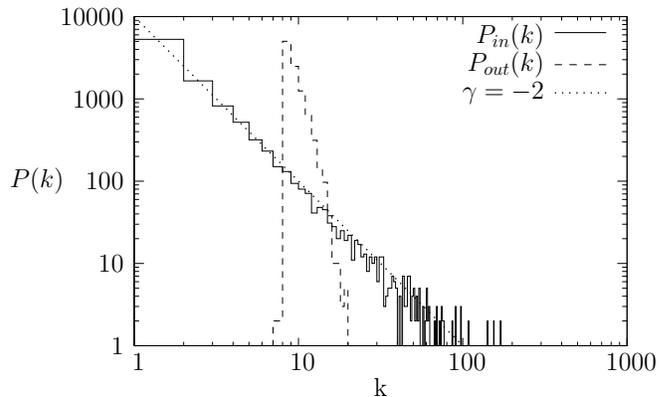}
\end{center}
\caption{\label{Pk} Distribution of input and output connections for
the scale-free network used in simulations. }
\end{figure}
Our procedure allows to generate a network that has a power-law
distribution of input contacts,  $P_\text{in}(k)\simeq k^{-\gamma}$,
with $\gamma \simeq 2$ (see Figure~\ref{Pk}), while the distribution
of output connections,
$P_\text{out}(k)$, is found to be exponentially distributed. This is
an
interesting feature of the model as the input connections represent
the total number of contacts to which an individual is exposed, while
the output connections represent the actively pursued contacts, e.g.
familiar ones. A customer, for
instance, is exposed to a large number of obliged contacts,
and may become infected with a large probability. These are
considered ``input'' links. On the other hand, people in a public
position is more monitored, and it is not plausible that they can
infect a comparable large number of people. Infection is limited to
the private sphere, where contacts are more intense. These are the
``output'' links.  We choose this algorithm
in order to have a
``worst-case'' scenario, with an exponent corresponding to a
diverging average of input
connectivity 

We have not studied the case of dynamic dependence of
the network on the infection level, however a high-level of
infection of a severe disease may surely induce changes in the
social network. It is reasonable to assume that, for mild diseases (or
diseases considered
harmless, like most of computer viruses), the social network is not
affected and only the level of prevention is
increased.

In the present paper we assume the effects of the infection
to be immediately visible, with no latency nor ``hidden infectivity''.
We also assume as temporal unit the time required to recover 
from illness without immunization and thus we explore the case of a
SIS dynamics.

\begin{figure}[t]
\begin{center}
\includegraphics[width=\columnwidth]
{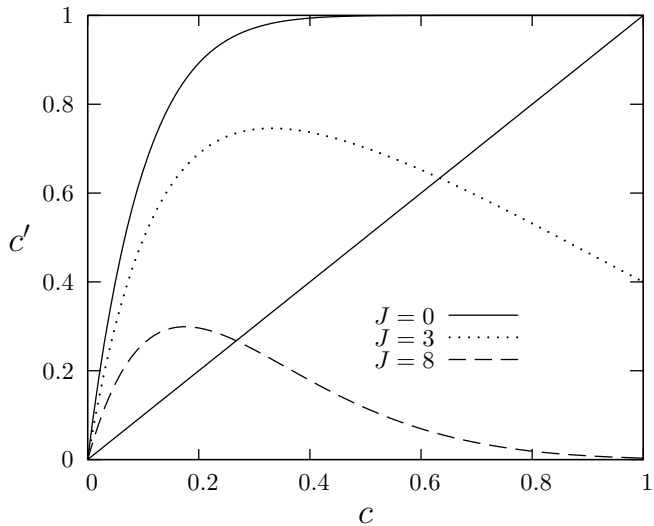}
\end{center}
\caption{\label{campomedio} Mean field return
map for fixed connectivity $z=10$, parameters $H=0$, $\alpha=1$,
$\tau=1$ and varying values of precaution level $J$. The effect of
risk perception ($J$) is to lower
the infectivity at high concentrations of infected individuals. }
\end{figure}
An individual can be infected separately by each of his neighbors with
a probability $\tau$ per unit of time (see Eq.~\eqref{cm}).

We model the effects of the perception of the risk of being infected
 replacing the bare infection probability  $\tau$ with $\tau I(s,k)$,
\begin{equation}\label{I} 
I(s,k)=\exp\left\{-\left[H+J\left(\frac{s}{k}\right)^\alpha\right]
\right\},
\end{equation}
where $k$ is the number of neighbors of a given site and $s$ is the
number of them that are ill. 

We assume the perception of the risk of being infected to depend
on the fraction of infected individuals among the neighbors, $s/k$, on
the level of precaution measures adopted, $J$, and on the use of
special prophylaxis, $\alpha \leq 1$. The quantity $H$ models a global
influence over the population, either  alarm of broadcasting media
news, in which case it could depend on the average level of the
infection. Its effect is that of reducing the bare infectivity
$\tau$, so in the following we only consider the case $H=0$. 
For the moment, we consider $\alpha=1$; the role of this
parameter will be clear in the following.  Differently from
Ref~\cite{Olinky}, in our model the
infectivity is not exclusively related to the connectivity.

The mean-field return map (for fixed connectivity $z$) is shown in
Figure~\ref{campomedio}. The effect of the introduction of risk
perception is evident: for high concentrations of infected individuals
the probability of being infected is diminished. Therefore, while for
$J=0$ and $z>1$ there is only one stable fixed point $c=1$ (all
individuals infected), by increasing $J$ one can have stable fixed
points $c<1$, limit cycles and even chaotic behavior~\cite{perpignan}.

\section{Mean field analysis}\label{meanfield}

The simplest mean-field approximation of the evolution of disease on a
network consists in neglecting correlations among variables. This is
essentially equivalent in considering the evolution on a tree, i.e.,
in assuming the absence of loops.

Let us denote with $c_k=c_k(t)$ the probability of having an infected
site of degree $k$ (with $k$ connections) at time $t$, and with
$c'_k=c_k(t+1)$ the probability at a subsequent time step.

The mean evolution of the system is generically given by 
\[
  c'_k = \sum_{\mathcal{C}_k} P_C(k|\mathcal{C}_k)
    P_I(k,\mathcal{C}_k) P_H(\mathcal{C}_k), 
\]
where $\mathcal{C}_k$ indicates the  local
configuration (degrees and
healthy status) at time $t$ around a site of degree $k$.
$P_H(\mathcal{C}_k)$ is
the probability of occurrence
of the healthy status of such configuration, $P_C(k|\mathcal{C}_k)$ is
the probability that the local configuration is connected to the site
under examination, and $P_I(k,\mathcal{C}_k)$ is the probability that
the disease propagates in one time step from $\mathcal{C}_k$ to the
site. 

In our case, the local configuration is given by a set of $k$ nodes,
of degree $(n_1, n_2, \dots, n_k)$, and status $(s_1, s_2,
\dots, s_k$), where $s_i = 0$ (1) indicates that the site $i$ is
healthy (ill). Thus $\mathcal{C}_k = (n_i, s_i)_{i=1}^k$ and
$P_H(\mathcal{C}_k) = \prod_{i=1}^k c_{n_i}^{s_i}(1-c_{n_i})^{1-s_i}$
since we assume decorrelation among sites. 

$P_C(k|\mathcal{C}_k)$ depends on
the assortativity of the network. Let us define $P_L(n|k)$ as the
probability that a site of degree $k$ is attached to a link connected
to a site of degree $n$. $P_L(n|k)$ is computed in an existing
network as the number of links that connects sites of degree $n$ and
$k$, divided by the total number of links that are connected to sites
of degree $k$, and $\sum_n P_L(n|k)=1$. The detailed balance
condition gives $k P_L(n|k) P(k)
= n P_L(k|n) P(n)$. For non-assortative networks, $P_L(n|k)=\phi(n)$,
and summing over the detailed balance condition one gets $P_L(n|k)=n
P(n)/z$, where $z$ is the average number of links per node, $z=\sum_k
k P(k)$. Assuming again a decorrelated network, we have 
\[
  P_C(k|\mathcal{C}_k)=\prod_{i=1}^k P_L(n_i|k)=\prod_{i=1}^k
\dfrac{n_i P(n_i)}{z}
\]
for non-assortative networks.

$P_I(k,\mathcal{C}_k)$ is the infection probability. In the case
without risk perception, it is
\[
  P_I(k,\mathcal{C}_k) =
\left(1-\left(1-\tau\right)^s\right),
\]
where $s=\sum_i s_i$. The risk perception is modeled by replacing
$\tau$ with $\tau \exp(-Js/k)$, which makes the equations hard to be
managed analytically except in the limit of vanishing infection
probability $c_k\rightarrow 0$, for which only the case $s=1$ is
relevant. We shall consider this point later. 

Putting all together, one gets
\[
\begin{split}
c'_k &= \sum_{\stackrel{n_1, n_2,\dots,n_k}{s_1, s_2,\dots,s_k}}
\left(\prod_{i=1}^k
P_L(n_i|k)c_{n_i}^{s_i}(1-c_{n_i})^{1-s_i}\right)
\\
&\qquad\cdot\left(1-\prod_i\left(1-\tau\right)^{s_i}\right).
\end{split}
\]

Using the relation 
\[
\sum_{x_1,x_2,\dots,x_k} \prod_i f(x_i)
= \left(\sum_x f(x)\right)^k, 
\]
 we obtain after some simplifications
\[
  c'_k = 1-\left(1-\tau\sum_n c_n  P_L(n|k)\right)^k.
\]
This expression could be obtained directly by noticing that $1-c$ is
the
probability of not being ill, which corresponds to the combined
probability of not being infected by any of the $k$ neighbors.
Neglecting correlations, these are $k$ independent processes
(although they depend on $k$). Each of these process is 1 minus the
probability of being infected, which is the sum, over all possible
degree $n$ of the neighboring node, of the probability that it is ill
($c_n$) times the probability that is is connected to the node under
investigation, $P_L(n|k)$.

Let us denote by $\overline{c}$ the asymptotic value of $c(t)$. 
Assuming that the transition between the quiescent ($\overline{c}=0$)
and active ($\overline{c}>0$) is continuous, its boundary is given by 
the values of parameters for which $c'/c=1$ in the limit
$c\rightarrow 0$. In this limit
\[
  c'_k \simeq k\tau\sum_n c_n  P(n|k),
\]
and we can now consider the case with risk perception, with $\tau$
replaced by $\tau\exp(-J/k^\alpha)$.

In the case of non-assortative networks,
\[
  c_k(t+1) = k
\frac{\tau}{z}\exp\left(-\frac{J}{k^\alpha}\right)\sum_n
c_n(t) n P(n).
\]
Calling $a(t+1) = \sum_n c_n(t) n P(n)$ (that does not depend on $k$),
we have $c_k(t) = (k \tau)/z \exp\left(-J/k^\alpha\right) a(t)$ and
thus
\[
 c_k(t+1) = c_k(t)\frac{\tau}{z} \sum_n
\exp\left(-\frac{J}{n^\alpha}\right)n^2 P(n).
\]
The critical boundary is therefore given by 
\begin{equation} \label{Jc}
 \sum_k
\exp\left(-\frac{J_c}{k^\alpha}\right)k^2 P(k)=\frac{z}{\tau},
\end{equation}
from which one could obtain  $J_c$ as a function of $\tau$ (we
replaced $n$ by $k$ for consistency with the rest of the paper). In
the
case $J=0$ (no risk perception), the formula gives
\[
 \tau_c = \frac{\mu_2}{z}= \frac{\langle
k^2\rangle}{\langle k\rangle},
\]
which is a well-known result~\cite{Boguna,MN2002}.

In the
case of fixed connectivity, $P(k) = \delta_{kz}$, and for $\alpha=1$
\begin{equation}\label{Jcfixed}
 J_c = z \log(\tau z). 
\end{equation}
In the absence of perception ($J=0$) one has $\tau_c=1/z$.

\begin{figure}
\includegraphics[width=\columnwidth]
{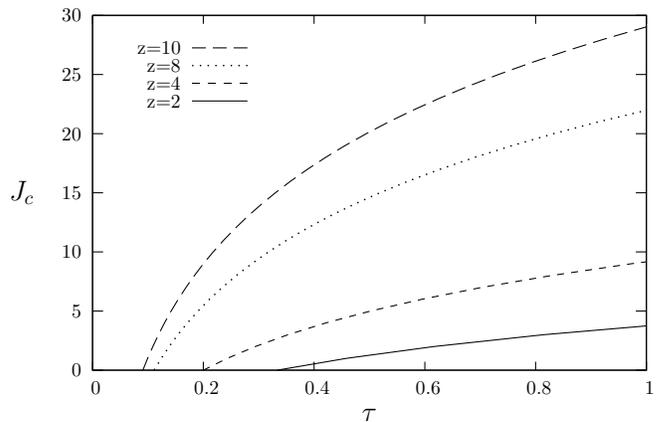}
\caption{\label{fig:poissonianJc} The mean-field dependence
of the critical value of precaution level $J_c$ with respect to the
bare infectivity $\tau$ for Poissonian networks, with average
connectivity $z$ and $\alpha=1$}
\end{figure}

For Poissonian networks (random graphs), 
\[
P(k) \simeq \frac{z^k e^{-z}}{k!}
\]
Numerical integration
of  Eq.~\eqref{Jc} for $\alpha=1$ gives the results shown in
Figure~\ref{fig:poissonianJc}. One can notice that for every value of
$\tau$ and finite average connectivity $z$, there is always a value
of the precaution level $J_c$ that leads to the extinction of the
epidemics. 

For non-assortative scale-free networks with exponent $\gamma$, $P(n)
\propto n^{-\gamma}$, the sum in Eq.~\eqref{Jc} diverges unless
$\gamma > 3$, irrespective of $\alpha$. 

This implies that at the mean-field level, any level of precaution is
not sufficient to extinguish the epidemics.

\section{Numerical results}\label{numerics}

The mean-field approximation disregards the effects of (correlated)
fluctuations in the real system. Indeed, the effects of random
and/or long-range  connections may disrupt correlations. We found
that the
behavior of microscopic simulations with random rewiring, both in the
quenched and annealed version, is well reproduced by
mean field simulations with a white noise term, with amplitude
proportional to $\sqrt{c\,(1-c)N}$. The noise term (or the
fluctuations in microscopic simulations) may bring the infection to
extinction if the average (or mean-field) oscillations come close to
$c=0$, as is often the case for a choice of $J$ for which chaotic
behavior appears in the mean-field approximation.
\begin{figure}[t]
\begin{center}
\includegraphics[width=\columnwidth]
{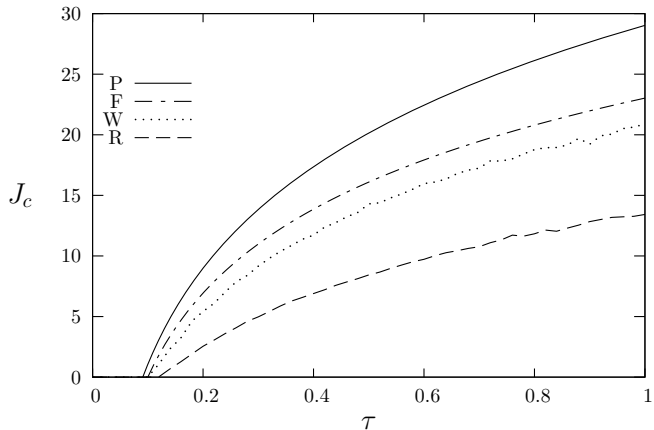}
\end{center}
\caption{\label{fig:Jc} Critical value $J_c$ of the precaution level
as a function of the base infectivity $\tau$
average connectivity
$k=10$ for the Poissonian mean-field (P), fixed connectivity mean
Field, Eq.~\eqref{Jcfixed} (F), and numerically ($N=1000$), for
the annealed rewired $p=1$ (W) and
regular one-dimensional (R) cases.}
\end{figure}

For regular
(fraction of rewired links $p=0$) and rewired ($p>0$) lattices, it
is always possible to observe a continuous transition towards a
critical level $J_c(\tau)$, such that the infection become
extincted, for every value of the bare infectivity $\tau$, as shown
in Figure~\ref{fig:Jc}.

For  scale-free networks, we concentrated on the case illustrated in
Section~\ref{model}, which can be considered a worst-case scenario
($\gamma=2$, diverging second and first moments of input
distribution). 

Simulations show that for $\alpha=1$ (Eq.~\eqref{I}), there
is no value of $J_c$ for which the infection may be stopped (although
not all population is always infected), for any value of $\tau$, in
agreement with the mean-field analysis.
\begin{figure}[t]
\begin{center}
\includegraphics[width=\columnwidth]
{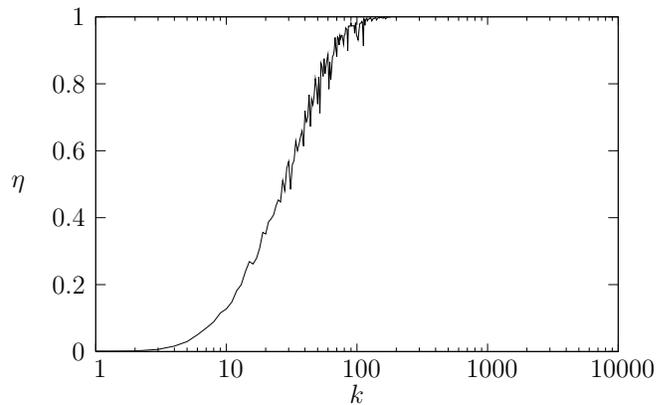}
\end{center}
\caption{\label{hubs} Fraction of time spent ill ($\eta$) in the
scale-free case, as a function of $k$ for  $K=10$, $J=10$. }
\end{figure}

The investigation of nodes that are \emph{more responsible} of the
spreading of the infection reveals, as expected, that the nodes with
higher input connectivity (hubs) stay ill most of time,
Figure~\ref{hubs}. Notice that also nodes with high input
connectivity have finite output connectivity, so the above relation
is not trivially related to the infection level.

\begin{figure}[t]
\begin{center}
\includegraphics[width=\columnwidth]
{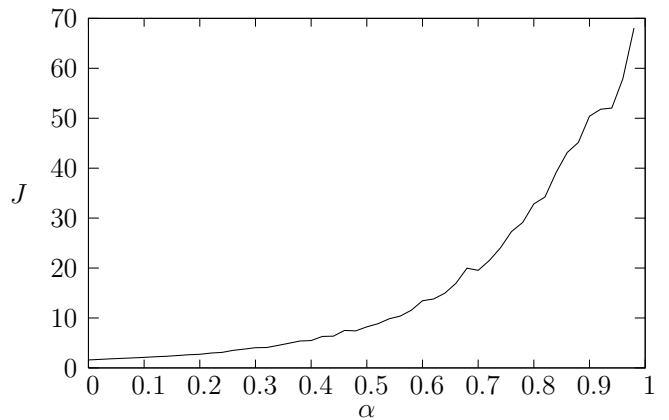}
\end{center}
\caption{\label{Jalpha} Dependence of the critical value of the
perception, $J_c$, as a function of the
\emph{exposure-enhanced} perception parameter $\alpha$, $K=4$,
$\tau=1$, $N=10000$. }
\end{figure}

In real life, however, public service workers who are exposed to
many contacts (like medical doctors, for instance) use additional
safety measures. In order to include this effect in the model, we
use the parameter $\alpha$, Eq.~\eqref{I}, that up to now have
been set to one. The effect of this parameter is to increase the
perception of the risk (or the safety measures) for nodes with higher
connectivity. As shown in Figure~\ref{Jalpha}, as soon as $\alpha <
1$, a finite critical
value of $J_c$ appears. The transition from the active ($c>0$) to the
absorbing ($c=0$) state occurs suddenly, due to fluctuations.
Essentially, nodes with high connectivity may fail to be infected due
to their increased perception of the infection, and this stops
efficiently the spreading. This effect is similar to targeted
immunization, but is not captured by the mean-field analysis. It
is a dynamical effect over a network far from the percolation
threshold, and thus containing loops.

The transition may be a finite-size effect,  related to the
unavoidable cut-off in
the degree distribution for finite populations, although simulations
with populations from $N=5000$ up to $N=80000$ do not show
any systematic change in the transition point.

\section{Conclusions}

In conclusion, we have studied  the effects of risk perception in a
simple SIS model for epidemics spreading. These effects
are modulated by two parameters, $J$ and $\alpha$, that 
reduce the infectivity of the disease as a function of the
fraction of people in the neighborhood that are manifestly ill. The
first parameter modulates the linear response, while the second
models non-linear effects like the increasing of prevention due to a
public exposed role. 
We found that for fixed or peaked connectivity there is always a
finite value $J_c$ of perception that makes the epidemics go
extinct. We studied the evolution of the disease in a ``worst case''
social networks, with scale-free input connectivity and
an exponent $\gamma\simeq 2$, for which both the average
input connectivity
and fluctuations diverge. In this case a
linear perception cannot stop the disease, but we found that, as 
soon as the perception is increased in a non-linear way ($\alpha<1$),
the epidemics may get extincted by increasing the perception level.
This latter transition is not continuous and is presumably induced
by fluctuations in hubs. It may be due to the finiteness of
population.

The mechanism that we propose is somehow analogous to vaccination of
hubs, except that is is a dynamics effect due to the local level of
diffusion of the disease, and is not exclusively related to local
connectivity. We think that a similar mechanism is at the basis of
the robustness of human population with respect to epidemics, even in
the absence of immunization procedures. One may speculate if, in
consequence of such robustness, humans have been selected to
exhibit visual signs of the most common diseases, which certainly
does not favors the spreading of infective agents. Another common
symptom of an illness is the tendency to isolation, which again could
be the result of selection. 


\section*{Acknowledgements}

L.S. research is supported by the EMBO organization under the contract ASTF 12-2007.
Authors acknowledge fruitful discussions with F. Di Patti and A.
Guazzini.


\begin{thebibliography}{99}
\bibitem{note} In SIR dynamics the infective node cannot be
reinfected, so $k$ is replaced by
 $k-1$.
\end{thebibliography}
\end{document}